\documentclass[11pt]{article}

\usepackage[margin=1in]{geometry}
\usepackage{amsmath,amssymb,amsthm,bm}
\usepackage{array}
\usepackage{graphicx}
\usepackage{placeins}
\usepackage{float}
\usepackage{algorithm}
\usepackage{algpseudocode}
\usepackage{xr-hyper}
\usepackage[hidelinks]{hyperref}
\usepackage[numbers,sort&compress]{natbib}
\usepackage{xcolor}


\newcommand{\R}{\mathcal R}
\newcommand{\E}{\mathbb E}

\newcommand{\Yacc}{\mathcal Y_{\rm acc}}
\newcommand{\Diss}{\mathcal D}

\newcommand{\Tr}{\operatorname{Tr}}

\newcommand{\dd}{\mathrm d}

\newtheorem{theorem}{Theorem}
\newtheorem{corollary}{Corollary}

\title{Measurement-Access Risk Frontiers for Autonomous Scientific Control}
\author{Bo Peng\\
\small Integrated Discovery Sciences Directorate\\
\small Pacific Northwest National Laboratory, Richland, Washington 99354, USA\\
\small email: \texttt{peng398@pnnl.gov}}
\date{}

\begin{document}
\maketitle
\begin{abstract}
Rapidly scaling autonomous science is limited not only by algorithms, compute or data volume, but by which physical records a platform exposes before action. We formulate physically accessible decision-making (PADM) and a measurement-access risk frontier: the Bayes-optimal target risk minimized over records realizable under cost, bandwidth, latency, disturbance, memory and actuation constraints. The frontier gives a no-free-autonomy limit: automation cannot collapse decision uncertainty by computation alone; an optimal controller cannot remove target components absent from its record, and closing that gap requires expanded access, auditing, tolerated disturbance, slower operation or restricted deployment. In monitored feedback, displacement-only control remains exposed to a hidden switching force, whereas a finite-bandwidth cue recovers part of the missing projection before action. A chemistry-aware candidate-ranking audit with a 1000-target stress panel, Gaussian sensing, hidden-regime decisions and cost-aware/thermodynamic channel selection provide reproducible checks. PADM identifies target-specific audit value and residual oracle gaps before deployment.
\end{abstract}


Autonomous science is moving from isolated demonstrations toward research infrastructure: robotic laboratories, self-driving materials platforms, autonomous synthesis laboratories, adaptive microscopes, closed-loop quantum experiments and AI-assisted controllers increasingly combine measurement, inference, planning and feedback~\cite{king2009automation,hase2019next,stach2021autonomous,abolhasani2023rise,tom2024selfdriving,burger2020mobile,macleod2020selfdriving,kusne2020closedloop,szymanski2023autonomous,merchant2023scaling,sayrin2011realtime,murch2013observing,minev2019catch}. When a platform also chooses measurements, the practical question is what the apparatus can know before it acts.

The intellectual machinery usually starts after an observation has been specified. Bayesian decision theory defines optimal actions and risks for a given observation \cite{wald1950statistical,savage1954foundations,degroot1970optimal,berger1985statistical,raiffa1961applied}; Blackwell informativeness and value-of-information theory compare observations by decision value \cite{blackwell1951comparison,blackwell1953equivalent,lindley1956measure,howard1966information}; and experimental design, Bayesian optimization, active learning, sensor selection, POMDPs and filtering optimize experiments or policies conditional on an observation model \cite{jones1998efficient,shahriari2016taking,kusne2020closedloop,macleod2020selfdriving,astrom1965optimal,smallwood1973optimal,kaelbling1998planning,bertsekas2012dynamic}. PADM asks the upstream architectural question: can the target-relevant record stream be generated before action?

In the laboratory, records are made by instruments with bandwidth, latency and memory limits; they can consume samples, disturb states, heat devices, induce photodamage, reset environments or alter the dynamics they reveal. Fisher information geometry describes local parameter directions \cite{amari2000methods,amari2016information}, while stochastic thermodynamics, Landauer--Bennett information thermodynamics and quantum measurement/filtering show that records can carry cost and back-action \cite{seifert2012stochastic,sagawa2010generalized,horowitz2014thermodynamics,parrondo2015thermodynamics,landauer1961irreversibility,bennett1982thermodynamics,wiseman2009quantum,jacobs2006straightforward,bouten2007introduction}. The missing question is concrete: which target-relevant variables can this platform expose under its physical constraints?

Current investment makes that question timely. National efforts such as the U.S. Department of Energy's Genesis Mission (\url{https://genesis.energy.gov/}) and autonomous-lab programs connect computing, artificial intelligence, facilities, data, synthesis, characterization and decisions. Before such platforms are scaled, their measurement architectures should be tested against task-required variables, consistent with platform- and task-specific self-driving-lab metrics \cite{volk2024performance,leong2025steering}. Manual wet-lab judgment often includes observations never formalized as variables, such as color, precipitate texture, humidity exposure, reagent age, clogging or batch history. When procedures move to robots, the question is which intuitions become channels, which variables disappear, and which audits should precede deployment.

This access question also appears at the controller level. A controller can optimize only over variables made available by its sensors, couplings, bandwidth, memory and actuation loop. Accurate displacement sensing can miss a switching force, and a sophisticated acquisition function can optimize an assay that omits the latent feature controlling target performance. This is the measurement-access limit, or no-free-autonomy: computation alone cannot collapse decision uncertainty. When a platform acts before exposing a target-relevant variable, the missing component remains as irreducible uncertainty. It can be reduced only by exposing the missing record, paying cost, tolerating disturbance, slowing or staging the loop, or restricting the operating domain. To our knowledge, PADM is the first architecture-level Bayes-risk framework at this intersection: physical measurement access, autonomous scientific control, a risk frontier and pre-deployment audit of target-relevant records. The novelty is not the Bayes projection identity alone, but the frontier obtained over constrained measurement architectures.

\begin{figure}[!t]
\centering
\includegraphics[width=\textwidth]{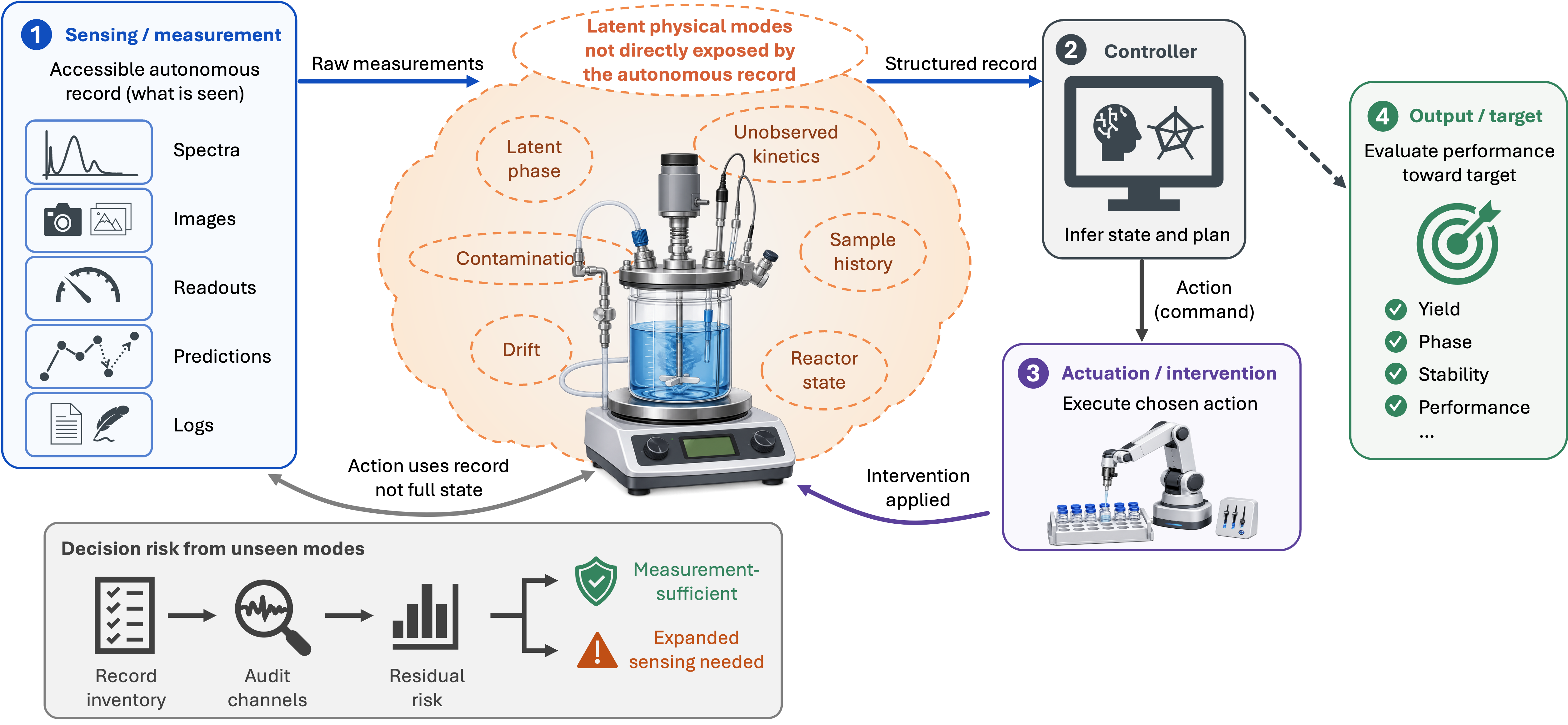}
\caption{\textbf{PADM procedure for auditing measurement access in autonomous scientific control.}
The numbered path separates the procedure into four parts. (1) The sensing/measurement block defines the autonomous record available before action, including spectra, images, scalar readouts, model predictions and experiment logs. (2) The controller uses only this record to infer the experimental state and plan the next intervention. (3) The actuation/intervention block executes the chosen action on the physical platform. (4) The output/target block evaluates the resulting performance against scientific targets such as yield, phase, stability and other task-specific metrics. The central reactor illustrates target-relevant latent physical modes--including latent phase, contamination, drift, unobserved kinetics, sample history and reactor state--that can affect the target without being directly exposed by the autonomous record. The lower audit branch is the PADM diagnostic: inventory the available records, identify candidate audit channels, estimate residual decision risk, and decide whether the architecture is measurement-sufficient or requires expanded sensing, an audit channel or a restricted autonomous operating domain.}
\label{fig:conceptual-geometry}
\end{figure}

With PADM, experimentalists ask what to log, audit, automate first, restrict or verify before autonomous control. Figure~\ref{fig:conceptual-geometry} separates the autonomous record, controller, intervention and target from latent modes outside the record, and frames PADM as an audit of whether expanded sensing, an audit channel or restricted autonomy is needed. The path below draws the frontier, makes it concrete in monitored feedback, states the risk-floor theorem and turns the projection identity into a workflow with benchmarks.

\section{Results}

\subsection{Measurement-access risk frontier}

The starting point is a physical observer-controller: an environment or simulator state \(S\), an observation architecture that generates a record \(Y\), a policy \(\pi\) that maps the available record to an action \(a=\pi(Y)\), and a loss \(\ell(a,S)\). The record may be a measurement, trajectory, counting record, monitored current, filtered estimate or experimental log. Microscopically, the observation architecture is a physical instrument: it generates a record and can disturb the subsequent state through back-action, sample consumption, heating, photodamage, reset, latency or actuator coupling. Let \(\Lambda=(C,B,\tau,\Delta_{\rm dist},M,U,\ldots)\) collect cost, bandwidth, latency, disturbance/back-action, memory, actuation and apparatus-specific limits. The feasible autonomous records, \(\Yacc^{\rm auto}(\Lambda)\), are the record-generating couplings and feedback loops realizable under \(\Lambda\). For a record \(Y\), let \(\mathcal P_{\rm adm}(Y)\) denote the admissible policies whose actions can depend only on \(Y\) before action. The autonomous risk frontier is
\begin{equation}
\R_{\rm auto}^*(\Lambda)
=
\inf_{Y\in \Yacc^{\rm auto}(\Lambda)}
\inf_{\pi\in \mathcal P_{\rm adm}(Y)}
\E[\ell(\pi(Y),S)] .
\label{eq:auto-risk-frontier}
\end{equation}
Here the expectation averages over the operating distribution of \(S\), the measurement noise that generates \(Y\), and any policy randomness. Thus \(\R_{\rm auto}^*(\Lambda)\) is the best risk attainable by any physically feasible autonomous record and any admissible controller using that record. It is not an algorithm benchmark, but a physical access bound: if a target-relevant variable is absent from every feasible autonomous record, no optimizer or learned policy can remove the corresponding residual risk. The record-set construction, expanded/oracle risks and measurement-sufficiency condition are collected in Supplementary Section 3.

Target risk enters through accessible decision-information geometry: the target-indexed map from physically realizable records to the target projections and residual risks those records induce. Each record \(Y\) defines an information subspace: the variables available before action. For a scalar decision target \(T=T(s)\), squared loss makes the geometry explicit. The Bayes-optimal action is \(\E[T|Y]\), the unresolved component \(T-\E[T|Y]\) gives the residual target risk, and an added channel is valuable exactly when it changes the target projection and reduces that residual variance. The audit workflow is not restricted to squared loss; for a general loss \(\ell(a,s)\), the same frontier is obtained by comparing Bayes risks for the admissible records. Thus the geometry records not just which data can be acquired, but which target-relevant directions those data expose before action.

Target-risk comparisons then become design diagnostics. An expanded architecture may add records, relax constraints or introduce complementary channels; its feasible set is \(\Yacc^{\rm exp}(\Lambda')\). Oracle access \(Y_{\rm or}\) is an ideal reference exposing the target-relevant variables needed to attain oracle risk. The corresponding expanded and oracle frontiers define physical automation gaps, $\Gamma_{\rm exp}(\Lambda,\Lambda') = \R_{\rm auto}^*(\Lambda)-\R_{\rm exp}^*(\Lambda')$ and $\Gamma_{\rm or}(\Lambda) = \R_{\rm auto}^*(\Lambda)-\R_{\rm or}^*$. These gaps measure inaccessibility rather than rank human and autonomous operators. For a specified target, loss and oracle reference, an architecture is measurement-sufficient at tolerance \(\epsilon\) when $\Gamma_{\rm or}(\Lambda)\le \epsilon$. The tolerance is task-specific: acceptable yield loss, energy error, defect density, safety margin or control regret. If the target depends on latent modes outside the record, the gap warns that additional sensing, slower operation or a restricted autonomous domain may be required. Candidate expanded channels are evaluated by target-specific risk reduction, and the remaining oracle gap determines whether the platform is measurement-sufficient.


\subsection{Monitored feedback as the central physical example}

\begin{figure}[!t]
\centering
\includegraphics[width=\textwidth]{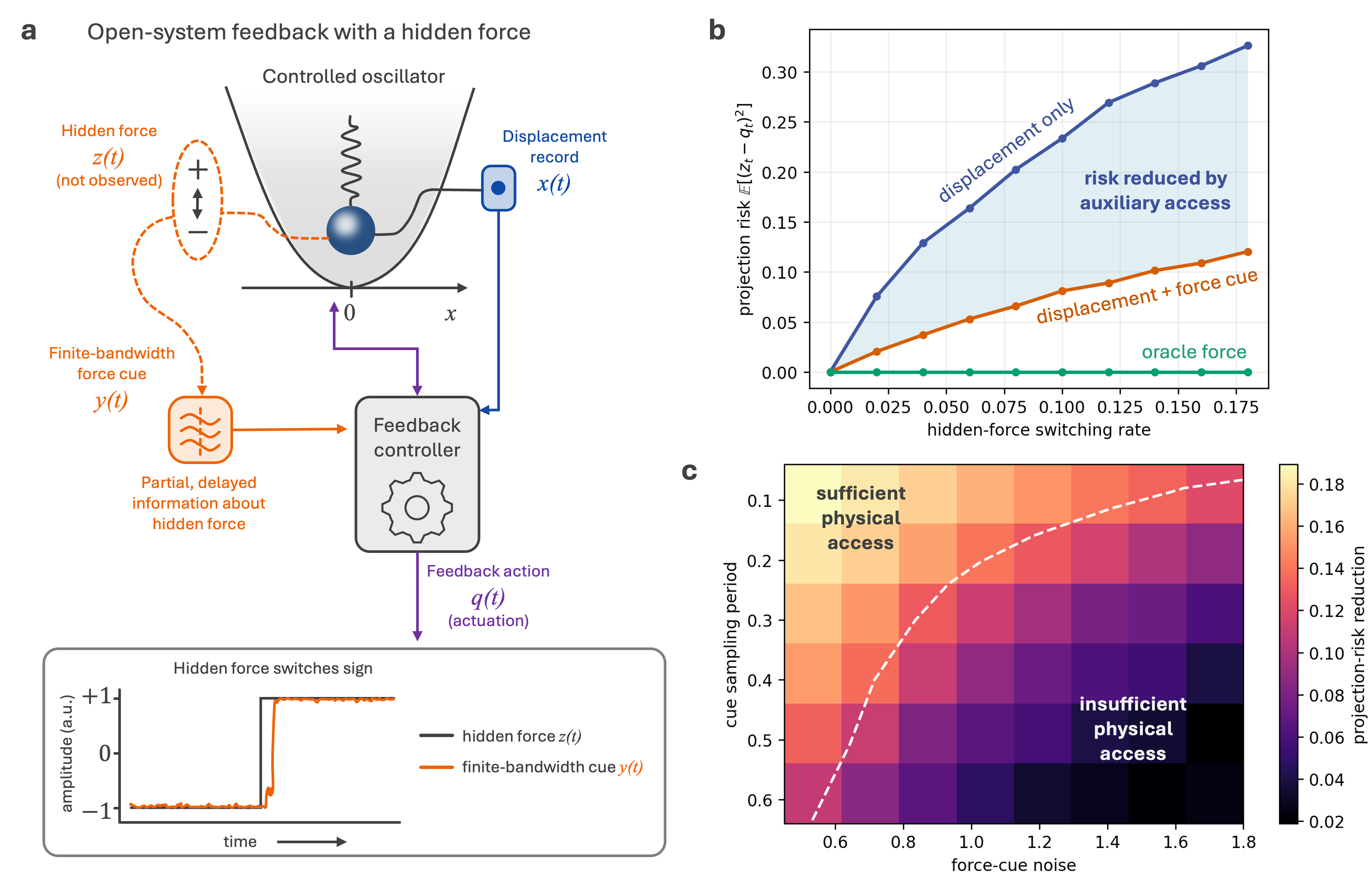}
\caption{\textbf{Open-system feedback is limited by finite-bandwidth access to hidden modes.}
A feedback controller observes a displacement record while a hidden environmental force switches sign.
\textbf{(a)} Controlled oscillator, displacement record, hidden force and finite-bandwidth auxiliary cue. The lower trace illustrates that the cue \(y(t)\) enters the controller record only after sampling or filtering, rather than as instantaneous access to \(z(t)\).
\textbf{(b)} Projection risk \(\E[(z_t-q_t)^2]\) versus hidden-force switching rate for displacement-only, displacement-plus-cue and oracle-force feedback. Faster switching exceeds the inference bandwidth of displacement-only feedback, while the auxiliary cue recovers part of the missing projection.
\textbf{(c)} Physical-accessibility map over cue noise and sampling period. Bright regions indicate larger target-specific risk reduction from the auxiliary channel; dark regions indicate insufficient access where cleaner or faster sensing, expanded sensing or slower autonomous operation is required. The dashed contour marks the illustrative sufficiency threshold \(\Delta R=0.10\).
Simulations integrate Eq.~\eqref{eq:main-open-system} with \(u_t=-fq_t\) and plot the residual risk in Eq.~\eqref{eq:main-dynamic-risk}. Default parameters are \(\Delta t=0.02\), \(k=1.5\), \(D=0.2\), \(f=1\), \(\gamma=0.08\), \(\sigma_r=0.85\) and \(m=8\); full grid sizes, burn-in and seeds are given in Supplementary Section 9. Independent-seed standard errors in \textbf{(b)} are smaller than the plotted markers.}
\label{fig:open-system-feedback}
\end{figure}

The monitored-feedback example shows the limit inside a live control loop. The autonomous record is the measured displacement trajectory, the missing target-relevant direction is a hidden environmental force, the complementary record is a finite-bandwidth force cue, and oracle access would reveal the force sign. To isolate timing, consider a minimal overdamped coordinate in a harmonic trap:
\begin{equation}
\dd x_t=(-kx_t+fz_t+u_t)\dd t+\sqrt{2D}\,\dd W_t,
\label{eq:main-open-system}
\end{equation}
where \(x_t\) is displacement, \(k\) stiffness, \(f\) hidden-force amplitude, \(z_t\in\{-1,+1\}\) the unobserved telegraph sign, \(u_t\) the applied control and \(D\) the diffusion strength. This Langevin testbed separates a visible coordinate from a target-relevant hidden mode. The expanded-channel record adds a noisy cue \(y_t\) sampled or filtered over an acquisition interval, not an instantaneous copy of the hidden force.

For an accessible record process \(Y\) generated by a measurement architecture, let \(\mathcal O_t^Y\) be the information retained by the controller before the feedback action at time \(t\). The key quantity is the record-conditioned estimate
\begin{equation}
q_t^Y=\E[z_t|\mathcal O_t^Y],
\qquad
u_t^Y=-fq_t^Y,
\qquad
\R_{\rm dyn}(Y)=\E[(z_t-q_t^Y)^2].
\label{eq:main-dynamic-risk}
\end{equation}
Here \(q_t^Y\in[-1,1]\) is the posterior mean for the current hidden-force sign. The feedback cancels the inferable force component, leaving residual force \(f(z_t-q_t^Y)\), so \(\R_{\rm dyn}(Y)\) is the normalized force-cancellation error. Later measurements can improve future estimates, but not the risk of an action already chosen from \(\mathcal O_t^Y\).

Figure~\ref{fig:open-system-feedback} connects record timing, dynamic error and design sufficiency. The cue enters the controller record only after sampling or filtering, so it is accessible information rather than oracle access (Figure~\ref{fig:open-system-feedback}a). As hidden switching accelerates, displacement-only inference loses the current force direction, while the finite-bandwidth cue recovers part of the missing projection before feedback is applied (Figure~\ref{fig:open-system-feedback}b). Sweeping cue noise and sampling period maps regimes where the platform is sufficient for this target and regimes requiring cleaner sensing, faster sampling or slower operation (Figure~\ref{fig:open-system-feedback}c). The useful variable is a timely record of the missing force direction.

The same residual has a direct physical consequence in the fixed-force limit. If \(z\) and the controller estimate \(q\) are held fixed over a relaxation interval, then the stationary displacement second moment is $\E[x^2|z,q]=\frac{D}{k}+\frac{f^2}{k^2}(z-q)^2$. With \(q\) fixed, incomplete access becomes a residual force and produces excess displacement variance. Averaging the excess term gives \((f^2/k^2)\R_{\rm dyn}(Y)\), so the same record \(Y\) sets the projection risk through \(q_t^Y\). Supplementary Section 8 connects that record to acquisition, storage, feedback and reset costs.

The record-generation point is not specific to this classical coordinate. For a controlled quantum density operator \(\rho\), the unconditioned dynamics can be written as
\begin{equation}
\dot\rho
=
-i[H(u_t),\rho]
+
\sum_\alpha \Diss[L_\alpha]\rho ,
\label{eq:main-lindblad}
\end{equation}
where \(H(u_t)\) is the control-dependent Hamiltonian and \(L_\alpha\) are jump or measurement operators. If only a subset of channels is monitored, the controller receives a stochastic record rather than the full state; in a diffusive quantum-trajectory representation, $\dd Y_\alpha(t) = \sqrt{\eta_\alpha}\, \Tr[(L_\alpha+L_\alpha^\dagger)\rho_c(t)]\,\dd t +
\dd W_\alpha(t)$, with detector efficiency \(\eta_\alpha\), conditional state \(\rho_c(t)\) and noise \(\dd W_\alpha\) \cite{breuer2002open,wiseman2009quantum,jacobs2006straightforward,bouten2007introduction}. These equations specify which channels generate accessible records and which remain unmonitored; back-action is part of the same instrument. Supplementary Section 7 gives the conditional evolution and accessible-set definition. For the theorem below, the feedback example supplies the target component \(z_t\), the available record \(\mathcal O_t^Y\), and the residual risk left outside that record.


\subsection{Measurement access and complementary channels}

The feedback example motivates a record-level statement. At the moment of action, the controller can use only the record it has already acquired; any target component outside that record remains as residual variance. The theorem below makes that projection statement independent of the oscillator, the hidden force and the particular cue.

\begin{theorem}[Autonomous-record risk floor]
Let \(T\in L^2\) be a scalar decision target and let \(Y_A\) be the record available to an autonomous controller before the action is chosen. For any square-integrable controller action \(a(Y_A)\), under quadratic loss,
\begin{equation}
\E[(T-a(Y_A))^2]
\geq
\E[(T-\E[T|Y_A])^2].
\label{eq:risk-floor}
\end{equation}
Equality is attained by the Bayes action \(a^*(Y_A)=\E[T|Y_A]\). Thus no controller using only \(Y_A\) can reduce target risk below the residual variance left after projecting \(T\) onto the autonomous record.
\end{theorem}

The theorem states no-free-autonomy conservatively. It grants the controller the optimal policy for its record, so the remaining risk is the target component not exposed by \(Y_A\). Automating before measuring such a component leaves irreducible uncertainty in the action: later outcomes can update future records, but not the risk of the action already chosen.

Mathematically, the proof mechanism is the orthogonal projection identity. Let \(Y_H\) be a candidate complementary record available before the same action. Write \(\R(Y_A)=\E[(T-\E[T|Y_A])^2]\) and \(\R(Y_A,Y_H)=\E[(T-\E[T|Y_A,Y_H])^2]\). Under quadratic loss, the target-specific value of adding \(Y_H\) is
\begin{equation}
\Delta_T(Y_H|Y_A)
=
\R(Y_A)-\R(Y_A,Y_H)
=
\E\!\left[
\left(
\E[T|Y_A,Y_H]-\E[T|Y_A]
\right)^2
\right]
\ge 0 .
\label{eq:main-projection-gain}
\end{equation}
Thus an added record is valuable exactly when it changes the Bayes-optimal projection of the target. A record can contain real environmental information and still have zero value for this target if the projection is unchanged; a noisy cue can be valuable if it resolves a missing latent target direction. The proof, conditional-variance form, equality condition and vector-target extension are in Supplementary Section 2.

\begin{corollary}[No-free-autonomy under record restriction]
Let \(\mathcal Y_1\subseteq\mathcal Y_2\) be feasible pre-action record families for the same target and loss. Then $\R^*(\mathcal Y_1)\geq \R^*(\mathcal Y_2)$. Restricting the pre-action record therefore cannot improve the best attainable Bayes risk. Automation is oracle-equivalent only when its feasible record family contains a measurement-sufficient record for the target; otherwise the residual oracle gap is a property of record access, not optimizer quality.
\end{corollary}

Together, Eqs.~\eqref{eq:risk-floor} and \eqref{eq:main-projection-gain} describe an information-access limit in the Bayes-optimal, known-model setting under quadratic loss and cost-free use of the added record. Finite data, model misspecification, learned representations, approximate inference and optimization error are separate computational limitations that can only increase realized risk relative to this physical projection bound, while costs, bandwidth, latency, disturbance, memory and actuation limits enter through \(\Yacc^{\rm auto}(\Lambda)\), \(\Yacc^{\rm exp}(\Lambda')\), or an explicit channel-selection objective.


\subsection{Operational audit of autonomous scientific control}

With the projection gain defined, the theorem becomes an audit of measurement sufficiency: does the closed loop record the variables that set the target before it acts? One fixes the target \(T\) or loss \(\ell(a,s)\), such as yield, phase purity, mobility, defect density, energy error or stability, and writes \(Y_A\) as the spectra, images, scalar readouts, logs, model predictions or sensor streams available before each action.

Candidate audit channels \(Y_j\) are evaluated by Eq.~\eqref{eq:main-projection-gain}: does \(Y_j\) change the target prediction beyond \(Y_A\)? The audit searches for target-moving latent modes absent from \(Y_A\), such as hydration, reagent aging, hidden phase, polymorph identity, drift, beam damage or thermal history. Candidate channels include spectroscopy, microscopy, destructive assay, replicate experiment, calibration standard, high-fidelity simulation or ex situ characterization, and become expanded control channels only if acquired before action with acceptable cost, latency and disturbance.

Algorithm~\ref{alg:padm-audit} summarizes the practical audit: define the target, inventory \(Y_A\), estimate autonomous and oracle risks, score candidate channels, and decide whether to add an expanded channel, slow or stage the loop, or restrict autonomy. Pilot data compare held-out risk for models trained with \(Y_A\) against \((Y_A,Y_j)\). These estimators inherit finite-sample, model-class and distribution-shift limits, but test whether the candidate channel reduces target-specific risk rather than merely adding data. The binary audit surrogate is documented in Supplementary Figure S3; a molecule-level known-target recovery audit is summarized in Figure~\ref{fig:molecule-audit-main} and validated in Supplementary Figure S4.

\begin{algorithm}[!t]
\caption{\textbf{PADM measurement-access audit for autonomous scientific control.}
The key score is target-specific risk recovery, not generic information content.}
\label{alg:padm-audit}
\small
\begin{algorithmic}[1]
\Require Target or loss \(T\) or \(\ell(a,s)\); autonomous pre-action record \(Y_A\); candidate audit or expanded records \(\{Y_j\}\); constraint vector \(\Lambda\); oracle proxy \(Y_{\rm or}\); sufficiency tolerance \(\epsilon\).
\Ensure Deployment decision: measurement-sufficient autonomy, audited/expanded-channel autonomy, slower or staged operation, changed target, or restricted autonomous operating domain.
\State Define the physical decision target, action timing, and loss used to evaluate the action.
\State Inventory only variables physically available before action: sensor streams, images, spectra, scalar readouts, logs, model predictions, calibration data, and retained memory.
\State Estimate \(\R_A\) from \(Y_A\), choose an oracle proxy \(Y_{\rm or}\), and compute the oracle gap \(\Gamma_{\rm or}=\R_A-\R_{\rm or}\).
\For{each feasible candidate channel \(Y_j\)}
    \State Estimate the target-specific audit gain
    \Statex \hspace{\algorithmicindent}\(\displaystyle G_j=\R(Y_A)-\R(Y_A,Y_j)\)
    \Statex \hspace{\algorithmicindent}analytically, by simulation, or from paired held-out pilot risks.
    \State Record the channel cost, latency, disturbance, bandwidth, memory burden, and actuation constraints.
    \State Compute the recovered oracle-gap fraction when the denominator is positive
    \Statex \hspace{\algorithmicindent}\(\displaystyle A_j=\frac{\R(Y_A)-\R(Y_A,Y_j)}{\R(Y_A)-\R(Y_{\rm or})}\)
    \Statex \hspace{\algorithmicindent}and skip this fraction if \(\R(Y_A)=\R(Y_{\rm or})\), since no oracle gap remains.
    \Statex \hspace{\algorithmicindent}When needed, also compute a platform-specific cost-aware score.
\EndFor
\If{\(\Gamma_{\rm or}\le \epsilon\)}
    \State Declare the architecture measurement-sufficient for this target.
\ElsIf{some feasible \(Y_j\) recovers enough oracle gap at acceptable burden}
    \State Use audited or expanded-channel autonomy.
\Else
    \State Add sensing, slow or stage the loop, change the target, or restrict the autonomous operating domain.
\EndIf
\end{algorithmic}
\end{algorithm}

The final audit decision uses the oracle gap and channel-specific recovery scores. Automation is measurement-sufficient at tolerance \(\epsilon\) when \(\Gamma_{\rm or}(\Lambda)\le \epsilon\). Staged or audited autonomy is preferred when a feasible channel recovers enough of the gap at acceptable cost. Example audit channels are summarized in Supplementary Table S1.

A chemistry-aware version is a candidate-ranking audit for de novo molecular-design workflows, tested by known-target recovery. A deterministic template generator proposes candidates from fixed seeds that exclude the withheld target; RDKit then performs sanitization, canonicalization, descriptor calculation and fingerprinting~\cite{rdkit}. The target is included only as a withheld/enumerated evaluation candidate, so the benchmark tests ranking and audit recovery rather than prospective molecular discovery. The autonomous record \(Y_A\) contains strings, descriptors, fingerprints, proposal history and a descriptor-profile surrogate score. The audit record \(Y_j\) contains chemistry-facing checks before acceptance: sanitization, single-component connectivity, valence/plausibility flags, forbidden-substructure filters, QED or synthetic-accessibility proxies and a batch/plausibility cue. The example remains a computational audit benchmark, not experimental chemical validation, because it tests whether \(Y_j\) changes target ranking and held-out recovery risk beyond \(Y_A\). In Figure~\ref{fig:molecule-audit-main}, descriptor-only ranking can favor an inserted, descriptor-matched phenol-plus-acetamide fragment artifact, whereas the audited record rejects that artifact and recovers the connected acetaminophen scaffold. Supplementary Figures S4 and S5 give backend checks and the \(1000\)-target adversarial control panel. Under the fixed acceptance rule, descriptor-only ranking selects the inserted artifact in \(1000/1000\) cases by construction, whereas the chemistry audit rejects all inserted artifacts, recovers \(998/1000\) targets at rank \(1\), and recovers \(1000/1000\) within the top \(5\). PADM treats plausibility, synthesis-route, forbidden-substructure, calibration-reaction, LC--MS/NMR and batch-state records as useful only if they change target ranking or held-out risk.

Figure~\ref{fig:molecule-audit-main} is therefore a pre-acceptance ranking-audit stress test, not a competitive generator, synthesis substitute or discovery benchmark. A deployed version would estimate the same score from paired generator outputs, batch logs, route checks, spectra and measured outcomes.

\begin{figure}[!t]
\centering
\includegraphics[width=0.72\textwidth]{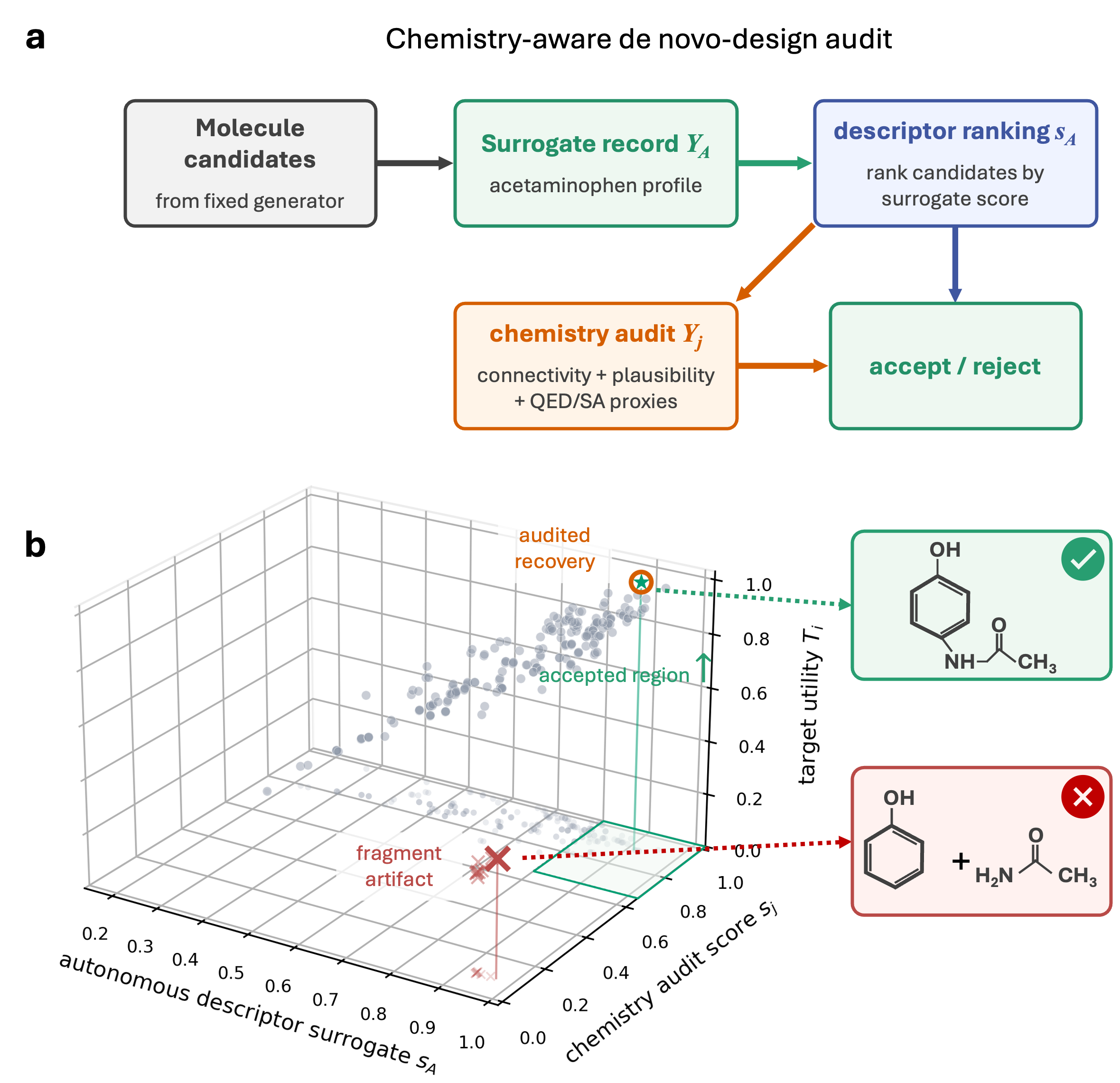}
\caption{\textbf{Chemistry-aware candidate-ranking audit for known-target recovery in de novo molecular-design workflows.}
\textbf{(a)} A deterministic template generator produces molecular strings and proposal history for the autonomous record \(Y_A\); a chemistry audit record \(Y_j\) adds RDKit-derived sanitization, single-component connectivity, valence/plausibility flags, forbidden-substructure checks, QED/synthetic-accessibility proxies and a pre-action plausibility cue before acceptance.
\textbf{(b)} Candidate decision map in autonomous surrogate score \(s_A\), audit score \(s_j\) and known-target utility \(T_i\), with a floor projection onto the \((s_A,s_j)\) plane. Gray points are ordinary generated candidates; faint red crosses show additional disconnected descriptor controls; the green star, large red cross and orange open circle mark the acetaminophen target, phenol-plus-acetamide artifact and audited selection, respectively. The shaded floor region indicates high-\(s_A\), high-\(s_j\) plausibility acceptance; final audited ranking also uses \(T_i\), so it need not choose the upper-right-most projected point.
The figure is a fixed-seed computational audit benchmark, not experimental chemical validation; backend reliability/regret, held-out audit-gain and chemistry-control checks are given in Supplementary Figure S4.}
\label{fig:molecule-audit-main}
\end{figure}


\subsection{Analytic benchmarks for access, hidden regimes and cost}

\begin{figure}[!t]
\centering
\includegraphics[width=\textwidth]{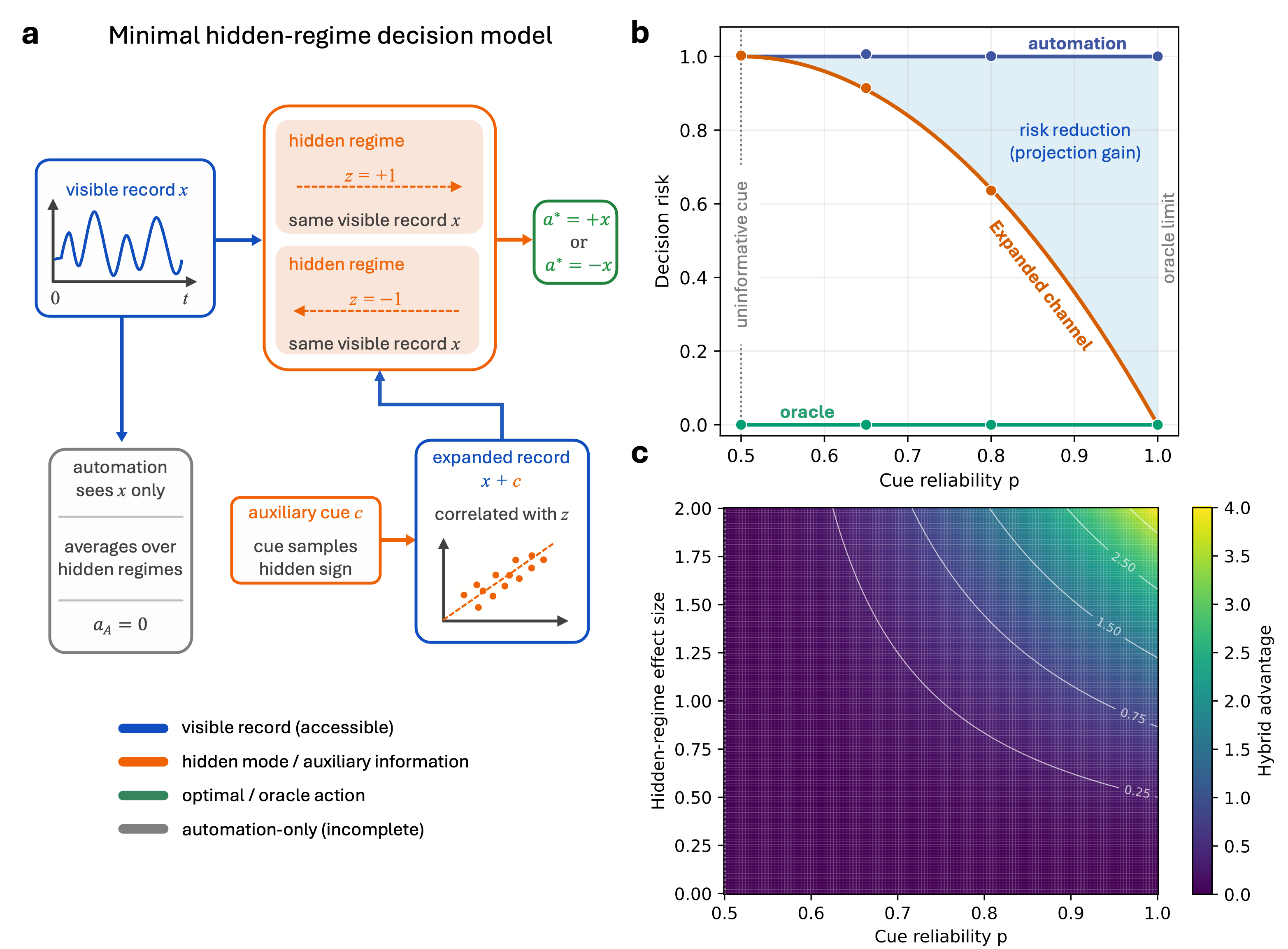}
\caption{\textbf{Minimal hidden-regime theorem benchmark for the decision risk floor.}
A visible record can be measured perfectly while still being insufficient for autonomous control if the target action depends on a hidden regime.
\textbf{(a)} Two hidden regimes have the same visible record \(x\), but opposite hidden signs \(z=\pm1\), which reverse the oracle action direction \(a^*=zx\). A controller observing only \(x\) averages over the hidden regime, whereas an auxiliary cue \(c\) with reliability \(p=\Pr(c=z)\) partially resolves the missing target-relevant mode.
\textbf{(b)} The automation-only risk has an irreducible floor, while the expanded-record risk decreases as the cue becomes informative and reaches the oracle limit when the hidden regime is resolved. The shaded region denotes the projection gain.
\textbf{(c)} Risk reduction appears only when the hidden regime affects the target and the auxiliary cue is correlated with that regime; it vanishes for an uninformative cue or for a target independent of the hidden mode.
Curves use \(x\sim\mathcal N(0,1)\), \(z=\pm1\) with equal probability, target \(T=\beta zx\) with \(\beta=1\), reliability grid \(p\in[0.5,1]\), and Monte Carlo validation points at \(p=0.5,0.65,0.8,1.0\) with \(1.2\times10^5\) samples per point.}
\label{fig:hidden-regime-demo}
\end{figure}

After the workflow examples, three minimal benchmarks isolate the audit mechanisms. The linear-Gaussian benchmark is the local calibration case: a Gaussian latent vector sets a linear target, the autonomous record observes one latent direction, a redundant record repeats it, and a complementary record observes an orthogonal direction. Gaussian conditioning makes the criterion transparent: a channel is useful only when its added precision overlaps an unresolved target direction. The posterior-precision formula, residual-risk expression and numerical checks are in Supplementary Sections 1 and 9.

Supplementary Figure S3 and Figure~\ref{fig:molecule-audit-main} turn the projection logic into autonomous-lab audit surrogates, with Supplementary Figure S4 reporting the molecule-audit backend checks. Figure~\ref{fig:hidden-regime-demo} isolates the minimal theorem benchmark: the visible variable is \(x\), the hidden regime is \(z\in\{-1,+1\}\), and the target is $T(x,z)=\beta zx$. The autonomous record \(Y_A=x\) can be perfect and still insufficient because the controller lacks the sign that determines the correct action, so \(a_A=\E[\beta zx|x]=0\). A complementary cue \(Y_H=c\), calibrated by \(p=\Pr(c=z)\), gives \(a_{A+H}=\beta(2p-1)cx\), while oracle access gives \(a_{\rm oracle}=\beta zx\). The corresponding risks are $\R_A=\beta^2\E[x^2]$, $\R_{A+H}=4p(1-p)\beta^2\E[x^2]$, and
$\R_{\rm oracle}=0$. Thus redundant access leaves the autonomous floor unchanged, complementary access reduces the projection risk when \(p>1/2\), and oracle access closes the gap. The point is access to the target-determining sign, not bandwidth alone. The derivation and numerical validation are given in Supplementary Sections 4 and 9.

The final benchmark adds cost. In the two-direction Gaussian model, autonomous access cheaply senses the visible direction but reaches the hidden direction only at high cost; a complementary architecture couples to both directions. The objective balances posterior-risk reduction against sensing cost, suppressing directions that are too expensive even if they contribute to risk. Supplementary Section 6 solves the convex problem. Supplementary Proposition S1 gives the binary-cue form: reducing hidden-direction risk by \(\Delta R=L_{\rm hid}(2p-1)^2\) requires cue mutual information \(I(Z;C)=\ln2-h_2[(1+\sqrt{\Delta R/L_{\rm hid}})/2]\), which lower-bounds target-linked cyclic memory/reset entropy production, up to implementation costs. Thus risk reduction requires target-relevant physical correlation, not free information.


\section{Discussion}

This work introduces a measurement-access risk frontier and audit workflow for autonomous-control measurement sufficiency. Its central message is no-free-autonomy: automation alone cannot collapse decision uncertainty. Autonomy can be oracle-equivalent only inside a measurement-sufficient operating domain; outside it, target components absent from the pre-action record remain irreducible. Closing that gap requires exposing the missing record, paying cost, tolerating disturbance, slowing or staging the loop, or restricting autonomy. Operationally, the audit defines the target, inventories pre-action records, estimates residual risk and decides whether the automation gap is acceptable under \(\Lambda\).

This framing separates the mathematical engine from the physical object. The Bayes projection identity is classical; the contribution here places it over a constrained measurement architecture and reads residual projection risk as a measurement-access floor. The identity is the proof engine, while the frontier is the object introduced here. Supplementary Section 7 derives accessible records from physical instruments, including quantum trajectories whose monitored channels also set back-action; Supplementary Sections 6 and 8 add cost and entropy accounting.

The comparison between autonomous, expanded-channel and redundant sensing is architectural. A fully autonomous controller is Bayes optimal when its record spans the target-relevant directions for the modeled environment and loss. An audit channel or complementary record helps only when it contributes a missing physical projection direction, reaches that direction at lower cost, or does so with less disturbance. A retrospective audit can diagnose a missing risk direction; it becomes an expanded control channel only if it can be acquired before the relevant action with acceptable cost, latency and disturbance. An exploratory public-record CAMEO/NIST materials audit applies the same held-out estimator to Fe--Ga--Pd records and finds that XRD structural summaries reduce magnetic-response prediction risk beyond a composition-only baseline under blocked splits and negative controls (Supplementary Figure S6).

The formulation is compatible with Bayesian experimental design, active learning, sensor selection, filtering, POMDPs and reinforcement learning, which optimize policies, experiments, sensors or beliefs conditional on an observation model or record stream. PADM asks whether that stream exposes target-relevant variables under physical constraints and whether a candidate expanded record reduces target-specific risk rather than merely adding data. Here, physical accessibility is treated at the level of generated records and decision risk. A natural next step is to derive admissible record sets directly from quantum instruments and monitored open-system dynamics, where changing the measurement channel also changes the system state through back-action, detector bandwidth, feedback delay, memory reset and entropy production. The result would be a risk--disturbance--cost frontier for autonomous feedback control.

PADM is therefore not a replacement for acquisition functions or policies; it is a preflight check on whether their input record can contain the variables the target actually requires. That check can be applied before deployment, during pilot campaigns or after logged runs reveal unexpected failure modes.

The bounds are known-model, Bayes-optimal measurement-access bounds for specified targets and losses. Finite-sample error, misspecification, learned representations, approximate inference and optimizer failure are separate computational limits that can only increase realized risk. Oracle access is a model reference, not a physically available controller. The demonstrations are analytic, synthetic or deterministic chemistry stress tests, chosen so projection risk, hidden-mode access, equality cases and cost bookkeeping can be checked directly. Thus the chemistry example remains an audit stress test, not validation of molecular synthesis. Deployment requires estimating the target, autonomous record, candidate audit channels such as LC--MS/NMR, batch logs or calibration reactions, feasible expanded records and oracle proxy. Pilot data can compare held-out risk for autonomous records alone versus autonomous plus candidate audit channels. This layer can combine with finite-sample learning theory, Bayesian experimental design, reinforcement learning and real autonomous-laboratory data without changing the bound.


\section*{Methods}

All numerical demonstrations use either synthetic data generated from the stated models or deterministic computational records generated by the stated chemistry-audit code, so that the simulated risks can be checked against closed-form Bayes-risk formulas or limiting cases. The purpose of the computations is therefore validation and illustration of the measurement-access bounds, not empirical discovery of the principle. The figure scripts regenerate the plotted main-text and supplementary figures, and unit tests check Gaussian conditioning, autonomous-lab audit formulas, chemistry-aware known-target recovery audit limits, hidden-regime analytic risks and equality cases, cost optimization, the binary memory-reset bound, monitored-feedback metrics and nonlinear multi-regime robustness. Detailed numerical settings, burn-in windows, sample counts, deterministic seeds and validation tests are given in Supplementary Section 9.

The monitored-feedback simulation integrates the stochastic hidden-force model in Eq.~\eqref{eq:main-open-system}. The autonomous and expanded-channel controllers both use the exact binary Bayes filter for the displacement record. The expanded-channel controller additionally receives a noisy finite-bandwidth force cue, while the oracle controller has direct access to the hidden force sign. The plotted dynamic risk is the post-burn-in residual projection risk \(\E[(z_t-q_t)^2]\), and the physical-accessibility map reports the risk reduction relative to displacement-only access. The filtering equations and open-system formulation are given in Supplementary Section 7.

The code also includes a retrospective-record audit runner for user-provided autonomous-science logs. The expected CSV schema records the target, autonomous pre-action columns, candidate audit-channel columns, optional oracle proxies and optional cost, latency or disturbance columns. The released code implements this protocol for user-provided CSV records and, in Supplementary Section 9.3, applies the same held-out estimator to public CAMEO/NIST records where target, autonomous record and candidate audit records are separable.

The autonomous-lab audit surrogates use binary hidden regimes to make the measurement-access calculation checkable. Supplementary Figure S3 uses the scalar target \(T=dz\) or optimum action \(a^*=dz\), with an autonomous record independent of the hidden regime and an audit cue with tunable reliability \(p=\Pr(c=z)\). Figure~\ref{fig:molecule-audit-main} uses the same audit logic as a chemistry-aware candidate-ranking audit for de novo molecular-design workflows, tested by known-target recovery: acetaminophen is excluded from the seed set and used as a withheld evaluation candidate; descriptor-only automation can rank a disconnected phenol-plus-acetamide artifact highly; and the RDKit-derived audit record supplies pre-acceptance connectivity, plausibility and QED/synthetic-accessibility proxy information. Supplementary Figure S4 provides the backend candidate-library checks, top-\(k\) recovery trace, reliability ablation, held-out audit-gain estimate and chemistry controls. The analytic risks, autonomy-gap recovery, cross-validated audit estimator and plotting parameters are given in Supplementary Section 9.

The Gaussian and hidden-regime benchmarks provide the two static analytic checks. In the Gaussian benchmark, a two-dimensional latent state is observed through one-dimensional noisy linear channels: the autonomous channel observes one direction, a redundant channel repeats that direction, and a complementary channel observes the unresolved direction. Posterior risks are computed from the Gaussian precision update and checked against Monte Carlo posterior-mean estimates. In the hidden-regime benchmark, independent samples of \(x\), \(z\) and \(c\) are used to compare the autonomous projection \(a_A=0\), the expanded-channel projection \(a_{A+H}=\beta(2p-1)cx\), and oracle access \(a_{\rm oracle}=\beta zx\). The Gaussian derivation is given in Supplementary Section 1, and the hidden-regime derivation is given in Supplementary Section 4.

The cost-aware benchmark uses a two-direction Gaussian decision geometry with linear directional sensing costs. The autonomous architecture has low cost for the visible direction and high cost for the hidden direction, while the expanded architecture can couple directly to both directions. The objective balances posterior trace risk against sensing cost, and the optimum follows from the separable convex solution. Supplementary Section 6 gives the analytic solution, parameter values and binary-cue memory-reset calculation.


\section*{Data availability}

Analytic, simulation and chemistry-stress-test data are generated by the released scripts. The shadow-mode public-record audit uses public CAMEO/NIST Fe--Ga--Pd records from Kusne et al. and the associated public repository/NIST data record. The CAMEO parser manifest, processed audit table, primary audit summary, robustness table, leakage/control report and scripts are included with the code.


\section*{Code availability}

Code for reproducing the simulations, figures and tests will be made available in a public repository upon publication. The code is available to editors and reviewers upon request during peer review. 


\section*{Acknowledgements}
This work was supported by the Early Career Research Program of the U.S. Department of Energy, Office of Science, under Grant No. FWP 83466.

\section*{Author contributions}
B.P. conceived the study, developed the theory, implemented the simulations, analysed the results, prepared the figures and wrote the manuscript.

\section*{Competing interests}
The author declares no competing interests.

\bibliographystyle{unsrtnat}
\bibliography{padm_refs}

@inproceedings{blackwell1951comparison,
  title={Comparison of Experiments},
  author={Blackwell, David},
  booktitle={Proceedings of the Second Berkeley Symposium on Mathematical Statistics and Probability},
  pages={93--102},
  publisher={University of California Press},
  year={1951}
}

@article{blackwell1953equivalent,
  title={Equivalent Comparisons of Experiments},
  author={Blackwell, David},
  journal={The Annals of Mathematical Statistics},
  volume={24},
  number={2},
  pages={265--272},
  year={1953}
}

@book{berger1985statistical,
  title={Statistical Decision Theory and Bayesian Analysis},
  author={Berger, James O.},
  publisher={Springer},
  year={1985}
}

@book{wald1950statistical,
  title={Statistical Decision Functions},
  author={Wald, Abraham},
  publisher={Wiley},
  year={1950}
}

@book{savage1954foundations,
  title={The Foundations of Statistics},
  author={Savage, Leonard J.},
  publisher={Wiley},
  year={1954}
}

@book{degroot1970optimal,
  title={Optimal Statistical Decisions},
  author={DeGroot, Morris H.},
  publisher={McGraw-Hill},
  year={1970}
}

@book{raiffa1961applied,
  title={Applied Statistical Decision Theory},
  author={Raiffa, Howard and Schlaifer, Robert},
  publisher={Harvard University Press},
  year={1961}
}

@article{lindley1956measure,
  title={On a Measure of the Information Provided by an Experiment},
  author={Lindley, Dennis V.},
  journal={The Annals of Mathematical Statistics},
  volume={27},
  number={4},
  pages={986--1005},
  year={1956}
}

@article{howard1966information,
  title={Information Value Theory},
  author={Howard, Ronald A.},
  journal={IEEE Transactions on Systems Science and Cybernetics},
  volume={2},
  number={1},
  pages={22--26},
  year={1966}
}

@book{amari2016information,
  title={Information Geometry and Its Applications},
  author={Amari, Shun-ichi},
  publisher={Springer},
  year={2016}
}

@book{amari2000methods,
  title={Methods of Information Geometry},
  author={Amari, Shun-ichi and Nagaoka, Hiroshi},
  publisher={American Mathematical Society and Oxford University Press},
  year={2000}
}

@book{wiseman2009quantum,
  title={Quantum Measurement and Control},
  author={Wiseman, Howard M. and Milburn, Gerard J.},
  publisher={Cambridge University Press},
  year={2009}
}

@book{breuer2002open,
  title={The Theory of Open Quantum Systems},
  author={Breuer, Heinz-Peter and Petruccione, Francesco},
  publisher={Oxford University Press},
  year={2002}
}

@article{bouten2007introduction,
  title={An Introduction to Quantum Filtering},
  author={Bouten, Luc and Van Handel, Ramon and James, Matthew R.},
  journal={SIAM Journal on Control and Optimization},
  volume={46},
  number={6},
  pages={2199--2241},
  year={2007}
}

@article{jacobs2006straightforward,
  title={A Straightforward Introduction to Continuous Quantum Measurement},
  author={Jacobs, Kurt and Steck, Daniel A.},
  journal={Contemporary Physics},
  volume={47},
  number={5},
  pages={279--303},
  year={2006}
}

@article{sagawa2010generalized,
  title={Generalized Jarzynski equality under nonequilibrium feedback control},
  author={Sagawa, Takahiro and Ueda, Masahito},
  journal={Physical Review Letters},
  volume={104},
  pages={090602},
  year={2010}
}

@article{parrondo2015thermodynamics,
  title={Thermodynamics of information},
  author={Parrondo, Juan M. R. and Horowitz, Jordan M. and Sagawa, Takahiro},
  journal={Nature Physics},
  volume={11},
  pages={131--139},
  year={2015}
}

@article{horowitz2014thermodynamics,
  title={Thermodynamics with Continuous Information Flow},
  author={Horowitz, Jordan M. and Esposito, Massimiliano},
  journal={Physical Review X},
  volume={4},
  pages={031015},
  year={2014}
}

@article{seifert2012stochastic,
  title={Stochastic Thermodynamics, Fluctuation Theorems and Molecular Machines},
  author={Seifert, Udo},
  journal={Reports on Progress in Physics},
  volume={75},
  number={12},
  pages={126001},
  year={2012}
}

@article{landauer1961irreversibility,
  title={Irreversibility and Heat Generation in the Computing Process},
  author={Landauer, Rolf},
  journal={IBM Journal of Research and Development},
  volume={5},
  number={3},
  pages={183--191},
  year={1961}
}

@article{bennett1982thermodynamics,
  title={The Thermodynamics of Computation--A Review},
  author={Bennett, Charles H.},
  journal={International Journal of Theoretical Physics},
  volume={21},
  pages={905--940},
  year={1982}
}

@article{astrom1965optimal,
  title={Optimal Control of Markov Processes with Incomplete State Information},
  author={{\AA}str{\"o}m, Karl J.},
  journal={Journal of Mathematical Analysis and Applications},
  volume={10},
  number={1},
  pages={174--205},
  year={1965}
}

@article{smallwood1973optimal,
  title={The Optimal Control of Partially Observable Markov Processes over a Finite Horizon},
  author={Smallwood, Richard D. and Sondik, Edward J.},
  journal={Operations Research},
  volume={21},
  number={5},
  pages={1071--1088},
  year={1973}
}

@article{kaelbling1998planning,
  title={Planning and Acting in Partially Observable Stochastic Domains},
  author={Kaelbling, Leslie Pack and Littman, Michael L. and Cassandra, Anthony R.},
  journal={Artificial Intelligence},
  volume={101},
  number={1--2},
  pages={99--134},
  year={1998}
}

@book{bertsekas2012dynamic,
  title={Dynamic Programming and Optimal Control},
  author={Bertsekas, Dimitri P.},
  publisher={Athena Scientific},
  year={2012}
}

@article{jones1998efficient,
  title={Efficient Global Optimization of Expensive Black-Box Functions},
  author={Jones, Donald R. and Schonlau, Matthias and Welch, William J.},
  journal={Journal of Global Optimization},
  volume={13},
  pages={455--492},
  year={1998}
}

@article{shahriari2016taking,
  title={Taking the Human Out of the Loop: A Review of Bayesian Optimization},
  author={Shahriari, Bobak and Swersky, Kevin and Wang, Ziyu and Adams, Ryan P. and de Freitas, Nando},
  journal={Proceedings of the IEEE},
  volume={104},
  number={1},
  pages={148--175},
  year={2016}
}

@article{king2009automation,
  title={The Automation of Science},
  author={King, Ross D. and Rowland, Jem and Oliver, Stephen G. and Young, Michael and Aubrey, Wayne and Byrne, Emma and Liakata, Maria and Markham, Magdalena and Pir, Pinar and Soldatova, Larisa N. and Sparkes, Andrew and Whelan, Kenneth E. and Clare, Amanda},
  journal={Science},
  volume={324},
  number={5923},
  pages={85--89},
  year={2009}
}

@article{hase2019next,
  title={Next-Generation Experimentation with Self-Driving Laboratories},
  author={H{\"a}se, Florian and Roch, Lo{\"\i}c M. and Aspuru-Guzik, Al{\'a}n},
  journal={Trends in Chemistry},
  volume={1},
  number={3},
  pages={282--291},
  year={2019},
  doi={10.1016/j.trechm.2019.02.007}
}

@misc{rdkit,
  title={RDKit: Open-source cheminformatics},
  author={Landrum, Greg and others},
  year={2025},
  howpublished={\url{https://www.rdkit.org}},
  note={Accessed for deterministic molecule parsing, descriptors and fingerprints}
}

@article{burger2020mobile,
  title={A Mobile Robotic Chemist},
  author={Burger, Benjamin and Maffettone, Phillip M. and Gusev, Vladimir V. and Aitchison, Catherine M. and Bai, Yang and Wang, Xiaoyan and Li, Xiaobo and Alston, Ben M. and Li, Buyi and Clowes, Rob and Rankin, Nicola and Harris, Brandon and Sprick, Reiner S. and Cooper, Andrew I.},
  journal={Nature},
  volume={583},
  pages={237--241},
  year={2020}
}

@article{macleod2020selfdriving,
  title={Self-Driving Laboratory for Accelerated Discovery of Thin-Film Materials},
  author={MacLeod, Benjamin P. and Parlane, Frederick G. L. and Morrissey, Timothy D. and H{\"a}se, Florian and Roch, Lo{\"\i}c M. and Dettelbach, Kevan E. and Moreira, Renato and Yunker, Lars P. E. and Rooney, Mary B. and Deeth, Jason R. and Lai, Vitaliy and Ng, Gillian J. and Situ, Helen and Zhang, Ruo and Elliott, Matthew S. and Haley, Thomas H. and Dvorak, David J. and Aspuru-Guzik, Al{\'a}n and Hein, Jason E. and Berlinguette, Curtis P.},
  journal={Science Advances},
  volume={6},
  pages={eaaz8867},
  year={2020}
}

@article{kusne2020closedloop,
  title={On-the-Fly Closed-Loop Materials Discovery via Bayesian Active Learning},
  author={Kusne, A. Gilad and Yu, Hongyu and Wu, Congyu and Zhang, Hongyi and Hattrick-Simpers, Jason and DeCost, Brian and Sarker, Sayantan and Oses, Corey and Toher, Cormac and Curtarolo, Stefano and Davydov, Albert V. and Agarwal, Ritesh and Bendersky, Leonid A. and Li, Mo and Mehta, Apurva and Takeuchi, Ichiro},
  journal={Nature Communications},
  volume={11},
  pages={5966},
  year={2020}
}

@article{szymanski2023autonomous,
  title={An Autonomous Laboratory for the Accelerated Synthesis of Inorganic Materials},
  author={Szymanski, Nathan J. and others},
  journal={Nature},
  volume={624},
  pages={86--91},
  year={2023},
  doi={10.1038/s41586-023-06734-w}
}

@article{merchant2023scaling,
  title={Scaling Deep Learning for Materials Discovery},
  author={Merchant, Amil and Batzner, Simon and Schoenholz, Samuel S. and Aykol, Muratahan and Cheon, Gregor and Cubuk, Ekin D. and others},
  journal={Nature},
  volume={624},
  pages={80--85},
  year={2023}
}

@article{sayrin2011realtime,
  title={Real-Time Quantum Feedback Prepares and Stabilizes Photon Number States},
  author={Sayrin, Cl{\'e}ment and Dotsenko, Igor and Zhou, Xingxing and Peaudecerf, Bruno and Rybarczyk, Th{\'e}o and Gleyzes, S{\'e}bastien and Rouchon, Pierre and Mirrahimi, Mazyar and Amini, Hadis and Brune, Michel and Raimond, Jean-Michel and Haroche, Serge},
  journal={Nature},
  volume={477},
  pages={73--77},
  year={2011}
}

@article{murch2013observing,
  title={Observing Single Quantum Trajectories of a Superconducting Quantum Bit},
  author={Murch, K. W. and Weber, S. J. and Macklin, C. and Siddiqi, I.},
  journal={Nature},
  volume={502},
  pages={211--214},
  year={2013}
}

@article{minev2019catch,
  title={To Catch and Reverse a Quantum Jump Mid-Flight},
  author={Minev, Zlatko K. and Mundhada, Shantanu O. and Shankar, Shyam and Reinhold, Philip and Guti{\'e}rrez-J{\'a}uregui, R. and Schoelkopf, R. J. and Mirrahimi, Mazyar and Carmichael, H. J. and Devoret, M. H.},
  journal={Nature},
  volume={570},
  pages={200--204},
  year={2019}
}

@article{tom2024selfdriving,
  title={Self-Driving Laboratories for Chemistry and Materials Science},
  author={Tom, Gary and Schmid, Stefan P. and Baird, Sterling G. and Cao, Yang and Darvish, Kourosh and Hao, Han and Lo, Stanley and Pablo-Garc{\'i}a, Sergio and Rajaonson, Ella M. and Skreta, Marta and Yoshikawa, Naruki and Corapi, Samantha and Akkoc, Gun Deniz and Strieth-Kalthoff, Felix and Seifrid, Martin and Aspuru-Guzik, Al{\'a}n},
  journal={Chemical Reviews},
  volume={124},
  number={16},
  pages={9633--9732},
  year={2024},
  doi={10.1021/acs.chemrev.4c00055}
}

@article{abolhasani2023rise,
  title={The Rise of Self-Driving Labs in Chemical and Materials Sciences},
  author={Abolhasani, Milad and Kumacheva, Eugenia},
  journal={Nature Synthesis},
  volume={2},
  pages={483--492},
  year={2023},
  doi={10.1038/s44160-022-00231-0}
}

@article{volk2024performance,
  title={Performance Metrics to Unleash the Power of Self-Driving Labs in Chemistry and Materials Science},
  author={Volk, Amanda A. and Abolhasani, Milad},
  journal={Nature Communications},
  volume={15},
  pages={1378},
  year={2024},
  doi={10.1038/s41467-024-45569-5}
}

@article{leong2025steering,
  title={Steering towards Safe Self-Driving Laboratories},
  author={Leong, Shi Xuan and Griesbach, Caleb E. and Zhang, Rui and others},
  journal={Nature Reviews Chemistry},
  volume={9},
  pages={707--722},
  year={2025},
  doi={10.1038/s41570-025-00747-x}
}

@article{stach2021autonomous,
  title={Autonomous experimentation systems for materials development: A community perspective},
  author={Stach, Eric and DeCost, Brian and Kusne, A. Gilad and Hattrick-Simpers, Jason and Brown, Keith A. and Reyes, Karina G. and Schrier, Joshua and Billinge, Simon and Buonassisi, Tonio and Foster, Ian and Gomes, Carla P. and Gregoire, John M. and Mehta, Apurva and Montoya, Joseph and Olivetti, Elsa and Park, Chanyeol and Rotenberg, Eli and Saikin, Semion K. and Smullin, Stephanie and Stanev, Valentin and Maruyama, Benji},
  journal={Matter},
  volume={4},
  number={9},
  pages={2702--2726},
  year={2021},
  doi={10.1016/j.matt.2021.06.036}
}

\end{document}